\documentclass[aps,preprint]{revtex4}
\usepackage{amsmath}
\usepackage{epsfig}

\input{tcilatex}

\begin{document}

\preprint{}
\title[Effective dimension of]{Effective dimension of quasiparticle states
and remnant Fermi surface in oxychlorides $Ca_{2}CuO_{2}Cl_{2}$ and $%
Sr_{2}CuO_{2}Cl_{2}$}
\author{V.A.Gavrichkov}
\affiliation{L.V.Kirensky Institute of Physics}
\pacs{PACS number}

\begin{abstract}
Formal similarity between $\overrightarrow{k}-$ area for the contribution
from ZR-states and the remnant Fermi surface has been found. Presence of the
contribution from $^{3}B_{1g}-$ states at other $\overrightarrow{k}-$ area
means the two-channel processes of scattering. As consequence, the method of
integration with a constant energy interval reproduces not only remnant
Fermi surface at least , but one of the contributions of above states to the
total spectral intensity
\end{abstract}

\volumeyear{year}
\volumenumber{number}
\issuenumber{number}
\eid{identifier}
\date[Date text]{date}
\received[Received text]{date}
\revised[Revised text]{date}
\accepted[Accepted text]{date}
\published[Published text]{date}
\startpage{101}
\endpage{102}
\maketitle
\tableofcontents

Oxychlorides $Ca_{2}CuO_{2}Cl_{2}$ and $Sr_{2}CuO_{2}Cl_{2}$ are
antiferromagnetic insulators with Neel temperatures of $255K$ and $247K$,
respectively. The spectral intensity in angle resolved photoemission
spectroscopy (ARPES) of\ the oxychlorides gives the information about the
nature of a single hole in $CuO_{2}$ - plane. Here, we are interested the
only the low-energy exitations or so-called first electron-removal states.
The spectral intensity of the first electron-removal state in the
oxychlorides -- amplitude quasiparticle peak (QP) shows a sudden drop along
a line in $\vec{k}$ - space which closely resembles the Fermi surface (FS)
obtained in the band-structure calculation or ARPES data of the optimally or
overdoped high temperature superconductors (HTSCs). Comparing the binding
energies of the first electron-removal states at $\vec{k}$ - wave vectors on
the remnant FS one can formally define a gap \cite{Remn.FS [1]}. The gap
dispersion of the QP peak along $\vec{k}$ - contour of the remnant FS is
close to the d- wave- superconductivity gap in the optimal and overdoped
HTSCs, i.e. there is a clear connection between all three energy gaps.

The recent ARPES -- experiments for the oxychlorides have shown that there
is a strong dependence of the amplitude of QP on a photon energy \cite%
{Matr.eff. [2]}. For $21eV$ - photon energy$~$ there is the clear QP at $%
\overrightarrow{k}=\left( \pi /2+\epsilon ,\pi /2+\epsilon \right) $ . And
the amplitudes of peak at $\overrightarrow{k}=\left( \pi /2-\epsilon ,\pi
/2-\epsilon \right) $ and $\overrightarrow{k}=\left( \pi /2+\epsilon ,\pi
/2+\epsilon \right) $ are approximately equal each other. In really, it is a
strange behavior of spectral intensity in a vicinity of the remnant FS where
we have the right to expect the asymmetry in the amplitudes of QP at these
two $\vec{k}$ - points irrespective of the photon energy. A probable
explanation is the matrix element effect in ARPES- data \cite{Matr.eff. [2]}%
. The later, in turn, points out that there is the different nature of the
first electron-removal state at the different $\vec{k}$ - wave vectors.
However, according to the generally accepted conclusion, these states have $%
A_{1g}$ - symmetry of Zhang-Rice (ZR-) state on all BZ. This statement is
coordinated with the polarized ARPES -- data \cite{Polar.ARPES [3]}. In
really, we need to classify the valence states according to the small
irreducible representations of $\ \vec{k}$ - group at the specific $\vec{k}$
- point. Therefore, the admissible symmetries for the first removal state
can be very different with regard for the nature of the partial contribution
to the total spectral intensity and $\vec{k}$ - point symmetry. The
generalized tight binding (GTB-) approach \cite{Red.to Hab.mod. [4]}
represents an interesting opportunity to investigate the reasons resulting
in the matrix element effects in ARPES- spectrum from above wider view.

Analyzing ARPES-data from the oxychlorides we deal with a charge-transfer
insulator \cite{Char.transf.ins. [5]} and the appropriate low-energy model
is reduced to the one-band Habbard model or extended $t-J$-model only in the
special case \cite{Red.to Hab.mod. [4]}. In this case there is the
nondegenerate ground singlet state. In HTSCs, as believe, the ground state
is the ZR-state. The last statement has basically the experimental character %
\cite{Remn.FS [1], Matr.eff. [2]}. In contrast, the theoretical
investigations don't give the unequivocal conclusion about a nature of the
ground state in HTSCs. In the ab-initio works \cite{Eto [6]} it was find
that ZR-state competes with $^{3}B_{1g}$ - triplet state, especially, in the
area of the doped HTSC-sructures. The calculations show the triplet level is
situated on $0.5-1eV$ above $A_{1g}$ - level \cite{Red.to Hab.mod. [4], B-A
[7]}. Moreover there is a crossover between the singlet and triplet levels
at some parameters of the calculations. The above magnitudes are
approximately equal to a width of the valence bands in HTSCs and we think,
that there is a problem to coordinate theoretical and recent experimental
conclusions.

In the works \cite{Red.to Hab.mod. [4],Our paper [8]} we have calculated the
dispersion, spectral intensity and parity of quasiparticle states along the
high symmetric directions of BZ, within the frameworks $d_{x}$ , $d_{z}$ , $%
p_{x}/p_{y}$ , $p_{z}$ - five orbital general tight binding (GTB-) approach.
The calculations reproduce the partial contributions from out-of-plane
states: $d_{z}$ and $p_{z}$ in $\Gamma $ - and $M$ - symmetric points of BZ,
due to a crossing and hybridization of the bands of ZR- and $^{3}B_{1g}$ -
valence states. It is surprising, that, with regard for the magnitudes of
partial contributions, a calculated picture of the spectral intensity and
its dependence on the polarization also are coordinated with observed ones %
\cite{Our paper [8]}.

In present work we reproduce the partial contributions from the in-plane and
out-of-plane states to the total spectral intensity on all BZ and give a
probable interpretation for the remnant FS in the oxychlorides as $\vec{k}$
-area for the in-plane contribution only. As well known, ZR- states are $%
A_{1g}$ - symmetrized combination of $3d_{x}$ , $2p_{x}/2p_{y}$- in-plane
orbitales and, in turn, the triplet states are $^{3}B_{1g}$ - symmetrized
combination of $3(2)p_{z}$ , $3d_{z}$, - out-of-plane orbitales on the
whole. Let's notice also, that the partial contribution from $2p_{x}/2p_{y}$
- in-plane orbitales to $^{3}B_{1g}$ - triplet states is not zero but is
small in the all BZ \cite{Our paper [8]}. We believed that in $%
\overrightarrow{k}$ -area for the basic contribution from the out-of-plane
states the dependence of the spectral intensity on the photon energy will
arise due to the possible three-dimensional effects (3D) (i.e. a monotonic $%
k_{z}$ -dependence). And in $\vec{k}$ - area for the basic contribution from
the in-plane states the oscillations of spectral intensity at the different
photon energies are possible only \cite{Matr.eff. [2]}. The oscillations are
caused by the diffraction of an photoelectronic wave on the periodic
structure of oxychlorides along c-axis. Namely in this mean we have used the
term: `` effective dimension'' placed in a title of the work.

We will start from the recent results obtained at $T=0K$ within the
framework of GTB- approach where alongside with $A_{1g}$ -state there is the 
$^{3}B_{1g}$ - state in the two-hole sector of Gilbert space. They may are
briefly summarized as the following:

- on the top of the valence band of oxychlorides in the antiferromagnetic
(AFM-) phase there is the pseudogap of the magnetic nature: $E_{S}\left( 
\overrightarrow{k}\right) \sim 0\div 0.4eV$ between the virtual level and
valence band. The pseudogap converts into zero at $\vec{k}=\bar{M}$ . The
virtual level has the small spectral intensity, proportional to the
concentration of the zero spin fluctuations $n_{0}$. The pseudogap
disappears in the paramagnetic (PM-) phase, where the dispersion of the
valence band is similar to the dispersion of the optimally and overdoped
HTSCs \cite{Wells [9]};

- the valence band of the oxychlorides has no only two-dimensional
character, and on BZ there are the $\vec{k}$ - areas where the triplet
contribution from $d_{z}$ - orbital of copper and $p_{z}$ - orbital of the
apical ions Cl or O isn't precisely equal to zero;

- the experimental k-dependences of QP-amplitude in the spectral intensity
of oxichlorides\cite{Wells [9]} are satisfactorily described by the $\vec{k}$
- dependences of the partial contribution to the total spectral density
obtained in the framework of GTB- approach. The feature of these
dependencies is the nonzero contribution from triplet states in $\Gamma $ -
and $M$ - symmetric points of BZ.

Thus, the presence of states with the symmetry of distinct from $A_{1g}$ -
symmetry, namely of triplet $^{3}B_{1g}$ - states, does not contradict the
ARPES-data. Further we, with a help of the similar approach, will reproduce
separately both the $A_{out}\left( k,E\right) $ -contribution from the
triplet out-of-plane states, and $A_{in}\left( k,E\right) $ -contribution
from the singlet in-plane states on all BZ. The details of GTB-approach and
magnitudes of parameters can be found in the works \cite{Red.to Hab.mod.
[4], par.GTBM1 [10], par.GTBM2 [11]}. The calculation was carried out in
AFM- and PM- phases at T=0K. In Fig.1,2(a,b,c) the amplitudes of the
quasiparticle peak for the partial contributions: $A_{in}\left( k,E\right) $
and $A_{out}\left( k,E\right) $ to the total spectral intensity are showed.
In Fig.1(d) the dispersion of the quasiparticle peak on all BZ is showed and
well reproduces one observed in \ \cite{Wells [9]}. In PM-phase (Fig.2(d))
the dispersion is similar to one in doped HTSCs. 
\begin{figure}[h]
\begin{center}
\epsfig{file=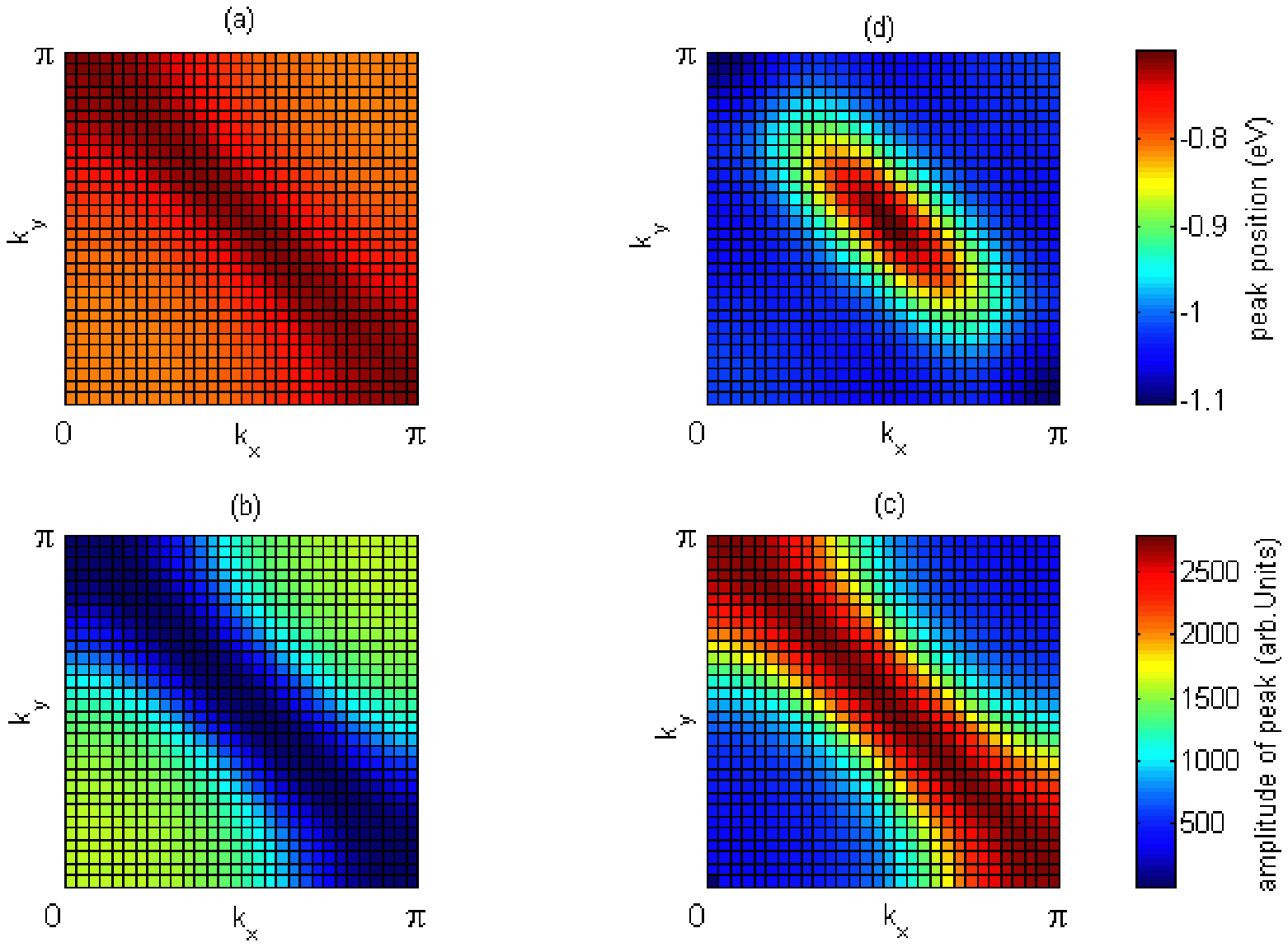,scale=0.4}
\end{center}
\par
\label{fig1abcd}
\end{figure}
\begin{figure}[h]
\begin{center}
\epsfig{file=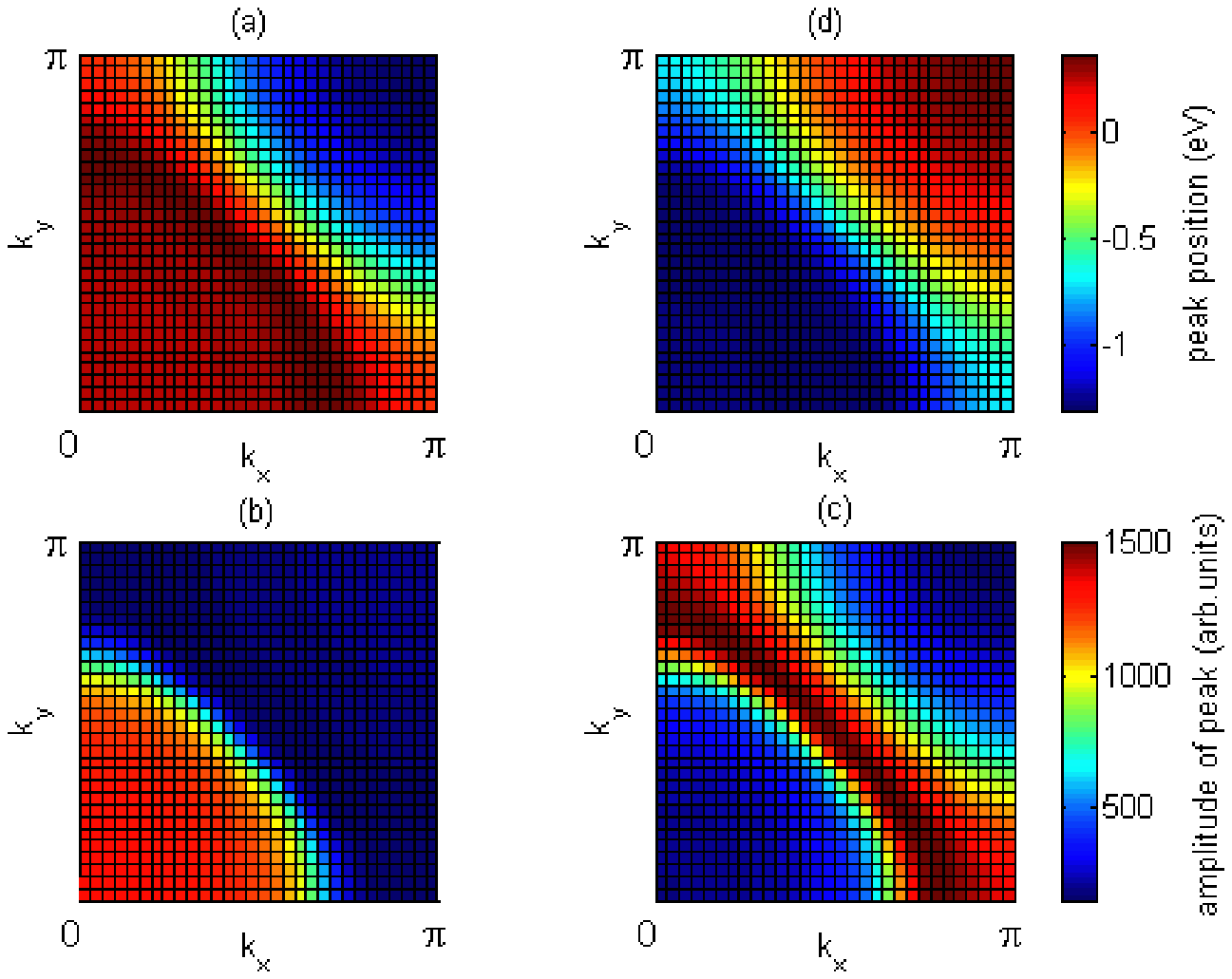,scale=0.4}
\end{center}
\par
\label{fig2abcd}
\end{figure}
We have found, that $A_{in}\left( k,E\right) $ - partial contribution is
distributed along $\vec{k}$ - contour close to the edge of the
antiferromagnetic BZ (Fig.1(c)) in AFM-phase. In PM-phase (Fig.2(c)), $\vec{k%
}$ - contour loses the symmetry and to become more close to one of the
remnant FS \cite{Remn.FS [1]}. We can suppose that it is not simple
coincidence provided that the procedure of recovery of FS by integration on
the fixed energy interval is insufficient in the considered case.

Here, both the dispersion, and spectral intensity are calculated without
taking into account the any processes of relaxation. There is only the
phenomenological damping $\epsilon =10^{-3}eV$ \ to obtain a lorentzian form
of curve of spectral intensity (Fig.3). 
\begin{figure}[h]
\begin{center}
\epsfig{file=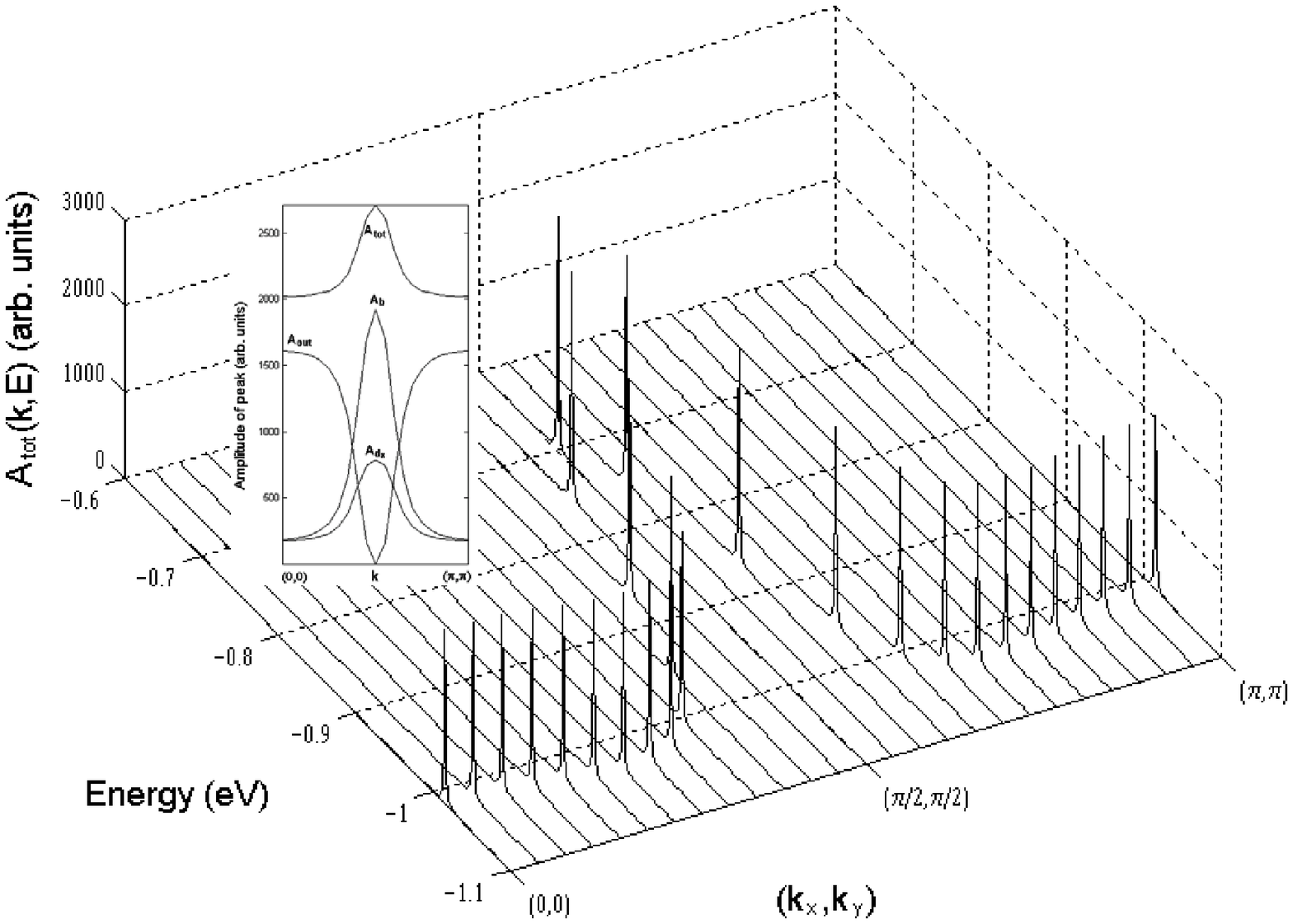,scale=0.4}
\end{center}
\par
\label{fig3}
\end{figure}
Specificity of considered materials is that namely the strongly correlated
nature of the quasiparticle states in BZ result in $\vec{k}$ - dependence of
the amplitude of QP. However, we think the specific singlet-triplet
character of the quasiparticle states means also the two-channel relaxation
process.

The calculations of an effective speed of relaxation with the account for
the strong correlated nature of the quasiparticle states within the
framework of the impurity diagram technique \cite{diagram [12]} show \cite%
{our diagram [13]}, that in according to Matissen's rule $\left( \tau
_{eff}^{\sigma }\right) ^{-1}=\frac{u_{\sigma }^{2}}{\tau _{in}}+\frac{%
\alpha _{\sigma }v_{\sigma }^{2}}{\tau _{out}}$ - the speed of relaxation is
additive function of the contributions with weights $u_{\sigma }^{2}$ and $%
v_{\sigma }^{2}$ , a sense of which consists in the probability with which
the carrier participates in the scattering on the appropriate potential for
in-plane and out-of-plane states. $u_{\sigma }^{2}+v_{\sigma }^{2}=1$ , and $%
\alpha _{\sigma }$ depends on the specific character of states and reflects
a presence of strong correlations renormalizing the hybridization effects of
the states. The scattering potentials for these groups of states can be
different, because their spatial nature is different also. It may be effects
nonstoichiometry along the in-plane and apical positions or fluctuations of
the magnetic moment in and between $CuO_{2}$ - planes. For example, we can
account the nonstoichiometry effects by means of the following arguments. In
the $\vec{k}$ - areas for only triplet contribution $v_{\sigma }^{2}\approx
1 $ and $\tau _{eff}^{\sigma }\approx \frac{\tau _{out}}{\alpha _{\sigma }}%
\sim ~c_{out}$ , where $c_{out}$ - concentration of the scattering centers
for carriers in the out-of-plane states. Along the $\vec{k}$ -contour of the
remnant FS accordingly $\tau _{eff}^{\sigma }\approx \tau _{in}~\sim c_{in}$
. Note also, that at $v_{\sigma }^{2}$, $u_{\sigma }^{2}\neq 0$ due to the
absence of the scattering in $CuO_{2}$ - planes $c_{in}\rightarrow 0$ it can
be obtained $\left( \tau _{eff}^{\sigma }\right) ^{-1}\approx \frac{%
v_{\sigma }^{2}}{\tau _{out}}$, i.e. the relaxation takes place even at
absence of scattering processes in $CuO_{2}$ - planes. Thus, crossing from
one of $\ \vec{k}$ - areas BZ in another one we should observe the widening
or narrowing of the quasiparticle peak in according to relation between $%
c_{out}$ and $c_{in}$ . In the case $c_{out}>>c_{in}$ the amplitude of
qusiparticle peak in $\vec{k}$ -area for the triplet states falls and to
collect all intensity the much greater energy interval than in $\vec{k}$
-area for contribution only from the in-plane states is required.

The two-channel scattering of carriers on the spin fluctuations may be more
actual in the oxychlorides. In the experiments using a magnetic neutron
scattering, in contrast the classical $K_{2}NiF_{4}$ where a transition to
the 3D- Bragg peak observed in 2\%- interval from $T_{N}$ , the dynamic
scattering in $La_{2}CuO_{4}$ transformed in the three-dimensional one from $%
T_{N}=195K$ gradually. Even at the high temperatures $\sim $ $350K$ the
low-energy spin fluctuations in $La_{1.89}Sr_{0.11}CuO_{4}$ have still $3D-$
character [13]. According to empirical conclusions, the lengths of spin
correlations are equal to $\xi _{||}\approx 18\pm 6$\AA\ and $\ \xi _{\bot
}\approx 5.5\pm 2.2$\AA . At the same temperature the carrier in the
out-of-plane state from $^{3}B_{1g}-$ symmetrized combination, situated at
the more hot spot of BZ than the carrier in the in-plane state from $A_{1g}-$
symmetrized combination.

Thus, the method of integration with a constant interval is unacceptable at
the presence of contributions to the total spectral intensity from so
different states. There is a risk, that the method of the integration used
in \cite{Remn.FS [1]} does not reproduce in particular not only the remnant
FS, but also one of the contributions to the total spectral intensity with
least relaxation time. Namely the contribution from $d_{x}$ -, $p_{x}$ / $%
p_{y}$ - orbitales.

We think, that as consequence of the scattering processes, the contribution
from the triplet states reproduces, in turn, a noncoherent part of the
spectral intensity in ARPES-spectra of oxychlorides. Therefore the effective
dimension of the quasiparticle states can rise up to 3D in the vicinity of $%
\ \Gamma $ - and $M$ - symmetric points of BZ. In these $\overrightarrow{k}$
- areas we can expect the monotonic dependence of the noncoherent part on
the photon energy. Probably, there is also an opportunity to observe not
only oscillations of the coherent part as it was made in work \cite{[14]},
but oscillations of the relation of the coherent and noncoherent parts of
the spectral intensity at the different photon energy.

\begin{acknowledgments}
We would like to thank S.G.Ovchinnikov for the discussion on this subject.
This work was supported by a Grant `` Enisey'' - 02-02-97705, INTAS
Prop.Number 654.
\end{acknowledgments}

\section{Capture of figures}

Fig.1 $\overrightarrow{k}$ - dependence of the: total spectral intensity
(a), out-of-plane spectral intensity (b), in-plane spectral intensity (c)
and QP position (d) in AFM at $T=0K$.

Fig.2 $\overrightarrow{k}$ - dependence of the: total spectral intensity
(a), out-of-plane spectral intensity (b), in-plane spectral intensity (c)
and QP position (d) in PM at $T=0K$.

Fig.3 3D-view on the quasiparticle peak in the total spectral intensity
along $\Gamma \leftrightarrow M$ - direction in AFM, $A_{tot}(%
\overrightarrow{k},E)=A_{d_{X}}(\overrightarrow{k},E)+A_{b}(\overrightarrow{k%
},E)+A_{out}(\overrightarrow{k},E).$ The inset shows $\vec{k}$ - dependence
of the partial contibutions to an amplitude of QP along $\Gamma
\leftrightarrow M$ - direction.


\begin{thebibliography}{99}
\bibitem{Remn.FS [1]} F.Ronning, C.Kim, D.F.Feng, D.S.Marshall, A.G.Loeser,
L.L.Miller, J.N.Eckstein, I.Bozovic, Z.X.Shen, Science, \textbf{282}, 2067
(1998).

\bibitem{Matr.eff. [2]} S.Haffner, D.M.Brammeier, C.G.Olson, L.L.Miller, and
Lynch, Phys.Rev. B \textbf{63}, 212501 (2001).

\bibitem{Polar.ARPES [3]} M.Grioni, H.Berger, S.Larosa, I.Vobornik, F.Zwick,
G.Mragaritondo, R.Kelley, J.Ma, M.Onellion, Physica B, \textbf{230-232}, 825
(1997).

\bibitem{Red.to Hab.mod. [4]} V.A.Gavrichkov, S.G.Ovchinnikov, A.A.Borisov,
and E.G.Goryachev, JETP, \textbf{91}, 2, 369 (2000).

\bibitem{Char.transf.ins. [5]} J.Zaanen, G.A.Sawatzky, J.W.Allen,
Phys.Rev.Lett. \textbf{55}, 4128 (1985).

\bibitem{Eto [6]} H.Kamimura, M.Eto, J.Phys.Soc.Jpn. \textbf{59}, 3053
(1990).

\bibitem{B-A [7]} H.Eskes, L.H.Tjeng, G.A.Sawatzky, Phys.Rev. B \textbf{41},
288 (1990).

\bibitem{Our paper [8]} V.A.Gavrichkov, A.A.Borisov, S.G.Ovchinnikov,
Phys.Rev. B \textbf{64}, 235124 (2001).

\bibitem{Wells [9]} B.O.Wells, Z.-X.Shen, A.Matsuura, et.al., Phys.Rev.Lett.%
\textbf{74}, 964 (1995).

\bibitem{par.GTBM1 [10]} L.F.Feiner, J.H.Jefferson, and R.Raimondi,
Phys.Rev. B \textbf{53}, 8751 (1996).

\bibitem{par.GTBM2 [11]} L.F.Feiner, J.H.Jefferson, and R.Raimondi,
Phys.Rev. B \textbf{53}, 8751 (1996).

\bibitem{diagram [12]} A.A.Abrikosov, L.P.Gor'kov, I.E.Dzyaloshinski,
Methods of quantum field theory in statistic physics, Gostehizdat, M, 350
pp. (1962).

\bibitem{our diagram [13]} V.A.Gavrichkov, S.G.Ovchinnikov, Fiz.Tverd.Tel. 
\textbf{41}, 68 (1999).

\bibitem{[14]} R.J.Birgeneau, G.Shiran, in $Physical$ $properties$ $of$ $%
High $ $Temperature$ $Superconductors$ $I$, edited by Donald M.Ginsberg
(World Scientific, Singapore, 1989), pp.151-211.

\bibitem{[15]} C.Durr, S.Legner, R.Hayn, S.V.Borisenko, Z.Hu, A.Theresiak,
M.Knupfer, M.S.Golden, and J.Fink, and F.Ronning, Z.-X.Shen, and H.Eisaki,
S.Uchida, and C.Janowitz, R.Muller, R.L.Johnson, and K.Rossnagel, L.Kipp,
G.Reichardt, Phys.Rev. B \textbf{63}, 014505 (2000).
\end{thebibliography}
\end{document}